\documentclass[12pt]{article}
\usepackage{graphicx}

\newcommand\pubnumber{SNSN-323-63}
\newcommand\pubdate{\today}

\def\infnpg{Istituto Nazionale di Fisica Nucleare,
Sezione di Perugia, I-06123 Perugia, ITALY}
\def\support{\footnote{Work supported by Cassa di Risparmio di Perugia.}}

\def\Title#1{\begin{center} {\Large #1 } \end{center}}
\def\Author#1{\begin{center}{ \sc #1} \end{center}}
\def\Address#1{\begin{center}{ \it #1} \end{center}}

\newcommand\pubblock{\rightline{\begin{tabular}{l} \pubnumber\\
         \pubdate  \end{tabular}}}
\newenvironment{Abstract}{\begin{quotation}  }{\end{quotation}}
\newenvironment{Presented}{\begin{quotation} \begin{center} 
             PRESENTED AT\end{center}\bigskip 
      \begin{center}\begin{large}}{\end{large}\end{center} \end{quotation}}

\input econfmacros.tex

\begin{document}
\begin{titlepage}
\pubblock

\vfill
\Title{CP (and CPT) violation studies at the Super Flavour Factories}
\vfill
\Author{ Elisa Manoni\support}
\Address{\infnpg}
\vfill
\begin{Abstract}
In this talk we present the perspectives about measurements of CP and CPT violating quantities at future Super Flavour Factories. In particular we will focus
on the expected sensitivities reachable after 5 years of data taking with the SuperB detector: this dataset will allow to perform measurements 
on \B meson, \D meson and \mtau lepton systems with greatly enhanced precision over the current results.    
\end{Abstract}
\vfill
\begin{Presented}
$6^{th}~International~Workshop~on~the~CKM~Unitary~triangle$\\
University of Warwick, UK,  September 6--10, 2010
\end{Presented}
\vfill
\end{titlepage}
\def\thefootnote{\fnsymbol{footnote}}
\setcounter{footnote}{0}

\section{Introduction}
The physics program of the B-factories, exploited in the last decade, has been extremely rich and successful
in confirming the Standard Model (SM) predictions in the flavour sector. A large part of the physics program has been devoted to 
CP violation (CPV) searches, mainly in the \Bd system but also in \D and \mtau decays. 
Comparing the current data with the theoretical expectations, a few discrepancies have emerged
and they need more precise measurements to be confirmed.
Accurate and redundant measurements of the same underlying quantity turn out to be crucial as also 
pointed out by studies which combine measurements from the flavour sector to test the CKM unitarity~\cite{UTFit},~\cite{CKMFit}.
One of the aims of the Super Flavour Factories, as SuperB, is to provide up to an order of magnitude improvement
in the Cabibbo-Kokayashi-Maskawa (CKM) matrix metrology precision.

\section{The SuperB project}
SuperB, together with Belle II~\cite{belle}, will be the next flavour factory generation running at the \FourS mass.
A detailed discussion on the SuperB physics program  and detector design can be found in~\cite{superb}.
SuperB will be an asymmetric \epem machine able to run at different energies 
in order to collect data at the $\Bd\bar{\B}_{d}$, $\Bs\bar{\B}_{s}$, and \ccbar  thresholds.
Moreover, the high $\epem \to \taup\taum$ cross section at the \FourS mass will allow to provide a high statistics \tautau pair sample.
The machine will be designed to reach an initial luminosity of $10^{36}~\mbox{cm}^{-2}~\mbox{s}^{-1}$ yielding $75~ab^{-1}$
in five years of data taking at the \FourS mass.\\
Given the similarities between BaBar and SuperB, to estimate the expected sensitivities in measurements of CP violation
after 5 years of SuperB, 
we have extrapolated the most recent BaBar results with some estimates on how the systematic errors
may improve.  In fact, some of them are statistical in origin, due to the finite size of data and Monte Carlo (MC) control samples, while
others require analysis technique improvements. In this study we will account only for sensitivi\-ty improvements related to the statistics increase,
benefits coming from detector upgrades and analysis technique refinements are currently under study.

\section{Measurements of $\beta$}
The $\beta$ CKM angle is extracted by a fit to the $\Delta t$
 distribution, $\Delta t$ being the difference between the two \B\footnote[2]{from here on, B means $B_{d}$ if not otherwise stated}'s decay times. 
 It is determined by the decay vertex of the fully reconstructed B in CP eigenstates ($\B_{sig}$) and 
 a second vertex, associated to the other B ($\B_{tag}$), computed by means of a multi-variate flavour tagging algorithm.
 Two crucial points in the $\Delta t$ computation will improve in SuperB with respect to BaBar. 
 The first is the resolution in the decay vertex determination: in SuperB this should benefit of reduced beamspot and 
 material budget and an additional inner layer in the silicon vertex detector.
 Also the performances of the flavor tagging algorithm should improve thanks to a larger tracking coverage, a better
particle identification system, and an improved vertexing.
All those effects are currently under investigation and for this talk the BaBar performances have been considered.\\
Up to now, \stwob has been measured with a 4\% precision in the $\b \to \ccbar \s$ golden mode~\cite{s2b_golden}. The main sources of 
systematic uncertainties are related to the determination of the parameters describing the $\Delta t$ resolution and the tagging performances, which makes use
of data control samples. The knowledge of the background properties accounts for 30\% of the total systematic error. Those factors can be reduced with higher 
statistics control samples. Moreover with larger datasets one can consider only the cleanest tagging categories (i.e. leptons) 
among the ones used in BaBar, and this would reduce the background amount. 
The increase in data statistics and the usage of the lepton tag only translate into an expected systematic error on \stwob of 0.005 (against the current 0.012).
The quantity $\stwob$ can be extracted also from modes such as $B^0\rightarrow\phi(\Kp\Km)\KS$ and $B \rightarrow \etapr\KS$ and the theoretical prediction
for the discrepancy between those and the golden mode measurement due to penguin pollution is $1\%-10\%$~\cite{s2b_peng}. 
The current knowledge of the CP parameter in the aforementioned modes has an uncertainty of 0.26  and 0.08 respectively and is statistical dominated. With $75~ab^{-1}$
statistical and systematic contributions will become comparable and the total error will reduce more than a factor 10 with respect to the current values.

\section{Measurements of $\alpha$}
The $\alpha$ angle is measured in $b\rightarrow u$ transitions mediated by both penguin and tree level diagrams.
Given the different weak phase of the two, the measured mixing angle $\alpha_{\mathit{eff}}$ 
differs from the CKM angle because of the penguin pollution ($\alpha_{\mathit{eff}}=\alpha-\Delta\alpha$). 
Exploiting isospin~\cite{a_su2} or flavor SU(3)~\cite{a_su3} symmetries, 
it is possible to disentangle penguin from tree contributions. One example is the SU(2) $\Bz \rightarrow \pi\pi$ analysis~\cite{a_ppbrr} where 6 physics 
observables are measured: $\Bz \rightarrow \pip\pim$, $\Bz \rightarrow \piz\piz$, and $\Bp \rightarrow \pip\piz$ branching fractions and $C(\pip\pim)$, $S(\pip\pim)$, and 
$C(\piz\piz)$ CP parameters. Also $S(\piz\piz)$ would play a role but it would require a good determination of the $\piz\piz$ vertex, not feasible
in BaBar. The preferred $\alpha$ solution for this measurement is $\alpha \in [71^{\circ},109^{\circ}]$ at 68\% confidence level. With $75~ab^{-1}$ of
SuperB data, it will be possible to measure  $S(\piz\piz)$ using photon conversion~\cite{a_pc} and further resolve ambiguites. With such statistics, 
$\alpha$ will be measured with a precision of the $1^{\circ}$ level. Since systematic uncertainties are dominated by estimation on data-Monte Carlo discrepancy 
affecting the selection efficiency, analysis refinements will further improve the expected error.
A second example is the $\B \rightarrow \rho\rho$ analysis~\cite{a_rrbrr}: here $\alpha$ is measured with a $6.5^{\circ}$ precision and there has been a $3\sigma$
evidence for the $\Bz \rightarrow \rho^0 \rho^0$ decay. The higher SuperB statistics will improve the measurement of such branching fraction and the precision on 
the CKM angle is expected to reach $1^{\circ}$. Moreover at high luminosities the systematic contribution will dominate and analysis improvements will be required.

\section{Measurements of $\gamma$}
The CKM angle $\gamma$ can be measured in $\Bpm \rightarrow \D^{(*)0}K^{(*)\pm}$ which proceeds through the interference between a color-favoured
$b \rightarrow c$ transition and a color-suppressed $b \rightarrow u$ decay, choosing D final states that are accessible both to \Dz and \Dzb.
Since only processes with tree level diagrams at leading order are considered,  negligible new physics effects are expected and the $\gamma$ measurement
turn out to be a standard candle as precision measurement in the SM. The sensitivity on $\gamma$ is proportional to $1/r_{B}$, 
being $r_{B} = \left | \mathcal{A}(b \rightarrow u) \right | /  \left | \mathcal{A}(b \rightarrow c) \right |$: due to the smallness of $r_{B}$ and of the 
branching fraction involved, the measurement is experimentally challenging. Depending on the chosen \D final state three methods can be exploited:
Dalitz or GGSZ method~\cite{g_dalitz} using Cabibbo favoured 3-body final states, GLW method~\cite{g_glw} using CP eigenstates, and
ADS method~\cite{g_ads} using doubly Cabibbo suppressed modes. Details on the different analysis strategies have been discussed at this conference 
for example in~\cite{g_gm}.
In the Dalitz method, $\gamma$ is measured from the Dalitz plot distribution of the \Dz daughters. The most recent BaBar measurement~\cite{g_bbr} quotes 
$\gamma = (68 \pm 14_{stat} \pm 4_{syst} \pm 3_{theo})^{\circ}$. The main contributions to the systematic error are statistical in origin. Scaling them and 
considering the theoretical uncertainties related to the Dalitz plot model the expected precision on $\gamma$ is $2.8^{\circ}$.  
By combining Dalitz and GLW analysis, this value is pushed down to $2.5^{\circ}$ and to $1.7^{\circ}$ when considering the
ADS method also. The Dalitz approach gives anyway the most precise constraint on the CKM angle
and at higher statistics will be essentially limited by the error related to the $D \rightarrow \KS \pi \pi$ Dalitz model.
A model-independent approach will allow to overcome this limit~\cite{g_ads},~\cite{g_dalitz}
and it will be exploited both by SuperB and LHCb.
Another way, probably less powerful,  to reduce the uncertainty on $\gamma$ would be to combine additive multi-body final states 
(i.e. $\D \rightarrow \KS\kaon\pi, \KS\piz\pi\pi, \kaon\kaon\pi\pi$) to the ones currently in use 
($\D \rightarrow \KS\kaon\kaon, \KS\pi\pi$) reducing the overall $\gamma$ error from $1.7^{\circ}$ to $1^{\circ}$. 
While waiting for SuperB results, the $\gamma$ measurement will be improved by 
LHCb which predicts an accuracy of $2-3^{\circ}$ with an integrated luminosity of $10~fb^{-1}$~\cite{g_lhcb}.

\section{CPV and CPTV measurements in neutral $B_{d}$ and $\tau$ systems}
CPT violating observables in the B system are related to parameters defining the B mass eigenstate ($B_{H,L}$) in the flavour eigenstate base
($\Bz,~\Bzb$): $\left | B_{H,L} \right> = p \sqrt{1 \mp z} \left | \Bz \right>  \pm q \sqrt{1 \pm z} \left | \Bzb \right> $. 
The CPT symmetry is conserved when z = 0 while $|q|^2+|p|^2=1$ if CP and CPT  symmetries hold.
The BaBar measurements~\cite{b_cpv} are consistent with the SM expectation (no CPT violation and indirect CP effects of the order of $10^{-3}$)
and provide a $3.3 \times 10^{-3}$, $8.0 \times 10^{-3}$, and $4.4\times 10^{-3}$ uncertainty on  $\left | q/p \right |$, $\Im z$, and $\Delta \Gamma \times \Re z$,
respectively.
The systematic uncertainties can be classified in three groups: reducible by means of higher statistics control samples (dominant for all the measured parameter combination), related to the precision of the other measurements such as $\Delta m_{d}$ and $\tau_{d}$ and errors requiring improvement in the knowledge of the detector 
(both not negligible for z related quantities). It has been estimated that with $75~ab^{-1}$ the errors on $\Im z$, and $\Delta \Gamma \times \Re z$ will reduce to 
$0.6 \times 10^{-3}$ and $0.3 \times 10^{-3}$, respectively.\\
In the SM, CPV in the $\tau$ system is predicted to be negligible. 
CP violation in the $\tau \rightarrow \pi \KS \nu$ decay mode is quoted to be
$3.3\times 10^{-3}$ with a $2\%$ precision~\cite{tau_cpv} and is essentially due to CPV in neutral K system.   
According to a preliminary result presented by BaBar at the ICHEP10 conference, the charge asymmetry for the 
$\tau \rightarrow h \KS \nu \mbox{n} \piz$ decay mode 
(being $h= \kaon , \pi$, and n the number of reconstructed neutral pions) is $-0.10 \pm 0.21 \pm 0.22$, consistent with no 
CPV. Systematic errors are mainly due to the determination of the detector charge asymmetry and selection biases 
evaluated by means of control samples. A reduction of such contribution,  currently under study, is  expected at SuperB.
A CPT violation test has also been performed.  Such a violation can be claimed if the two quantities
$r_{m} = \frac{m_{\taum}-m_{\taup}}{m_{\taup}+m_{\taum}}$ and $r_{\tau} = \frac{\tau_{\taum}-\tau_{\taup}}{\tau_{\taup}+\tau_{\taum}}$ 
are not consistent with zero. They have been measured to be $r_{m}=( -3.4 \pm 1.3 \pm 0.3) \times 10^{-4}$~\cite{tau_m}
and  $r_{\tau}=( 0.12 \pm 0.32 )$~\cite{tau_tau} where the uncertainty in the latter is statistical only. The systematic contribution to
$r_{m}$ is mainly due to the knowledge of the magnetic field, crucial in the momentum determination, and to the support tube budget material 
and will require some challenging study to be reduced. As for $r_{\tau}$, SuperB aims to a $10^{-4}$ precision after 5 years of data taking.

\section{Conclusions}
\begin{figure}[htb]
\centering
\begin{tabular}{cc}
\multicolumn{2} {c}{\includegraphics[height=1.9in]{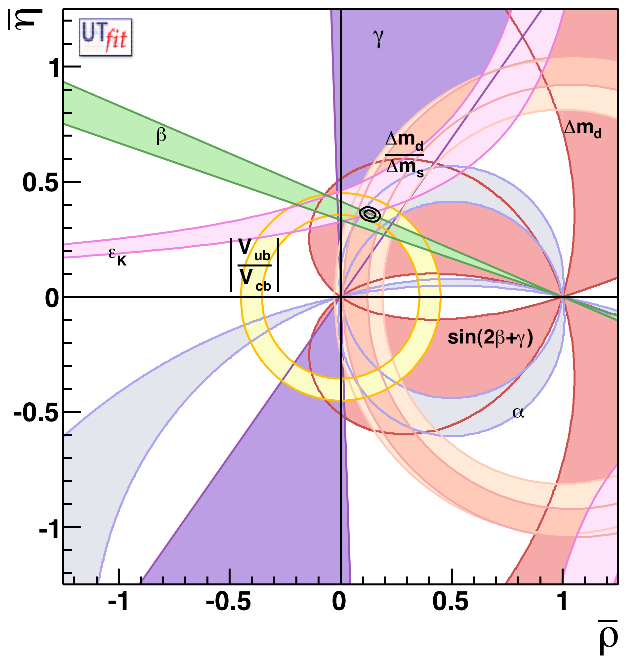}} \\
\includegraphics[height=1.8in]{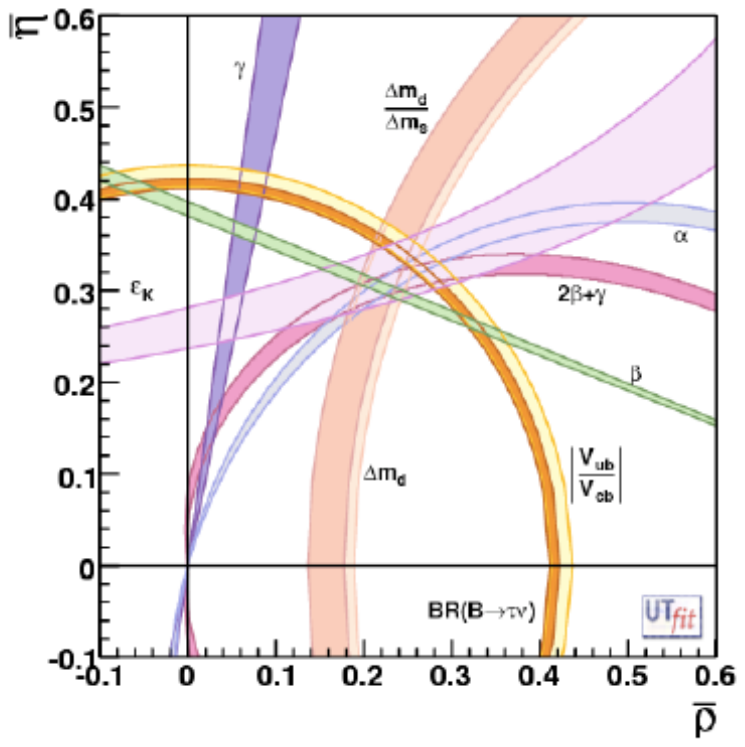} & 
\includegraphics[height=1.8in]{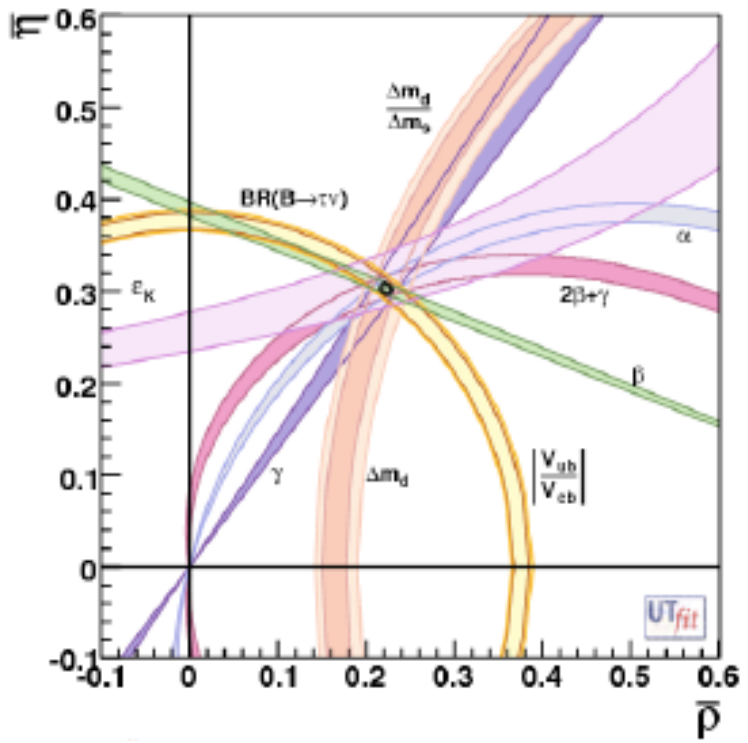} \\
\end{tabular}
\caption{Allowed regions for $\bar \rho$ and $\bar \eta$ from the UTFit analysis~\cite{UTFit}: before the ICHEP10 input updates (top),
with the expected lattice calculation available at Superb-$50~ab^{-1}$ time (bottom left), 
if SM holds with current theoretical calculation (bottom right). }
\label{fig:utfit}
\end{figure}

Figure~\ref{fig:utfit} shows the CKM fit results by the UTFit collaboration in three different scenarios:
the knowledge before ICHEP10 input updates (top); the "dream" that will show up if with $50~ab^{-1}$ the current tensions will be confirmed
and the constraints will not converge to a single point, indicating
that new physics modifies our understanding of quark
mixing (bottom left); the "nightmare" in which the SM prediction still holds (bottom right).
The last two examples, compared to the current result, 
show how a precise CKM metrology is still crucial to deeply understand the flavour sector of the SM.
Extrapolating the most recent BaBar results to $75~ab^{-1}$ integrated luminosity, it has been shown that
\stwob can be determined with a $5 \times 10^{-3}$ error in the golden channels; $\alpha$ and $\gamma$ will be both
known  with a precision of the degree level. Also the error on the CPV and mixing parameters in $\B$ system will be reduced 
by more than a factor ten while improvements on CPV measurement on the $\tau$ sector are currently under study.
The extrapolation includes only benefits from the increase in statistics and further improvement will come from 
 detector upgrades and analysis technique refinements. 
 
\end{document}